# Locations of Auroral Kilometric Radiation Bursts Inferred From Multi-Spacecraft Wideband Cluster VLBI Observations I: Description of Technique and Initial Results

R. L. Mutel[1], D. A. Gurnett[1], I. W. Christopher[1], J. Pickett[1], and M. Schlax[1]

**Abstract.** The Cluster Wideband Data instrument has been used to determine the locations of auroral kilometric radiation (AKR) using very long baseline interferometry. The technique involves cross-correlating individual AKR bursts from all six Cluster baselines using time and frequency filtered waveforms. The resulting differential delay peaks are used to determine source locations with an uncertainty as small as 500 km in a plane perpendicular to the source-spacecraft line of sight, or about 200 km when the burst position is projected onto the auroral zone along the magnetic field lines passing through the source. The uncertainty along the line of sight is much larger, but this is mitigated by assuming that the emission arises from a height corresponding to the electron gyro-frequency.

We report the locations of over 1,700 individual AKR bursts during six observing epochs between 10 July 2002 and 22 January 2003 when the Cluster constellation was high above the southern or northern hemisphere. In general we find that the AKR burst locations lie along magnetic field lines which map onto the nighttime auroral zone as expected from previous AKR studies. For the three observing epochs viewing the northern hemisphere, there is a strong tendency for AKR burst locations to be centered within the auroral oval and in the evening sector. The southern hemisphere burst locations favor magnetic local midnight to early morning and have somewhat higher invariant magnetic latitudes. The distribution of AKR auroral footprint locations at each epoch had a overall spatial scale between 1000 - 2000 km, much larger than the positional uncertainty of an individual AKR burst location magnetic footprint, but a small fraction of the auroral oval. This indicates that on a timescale of 1-3 hours conditions for suitable generation of AKR emission are found on only a fraction of all magnetic field lines crossing through the auroral oval. For two of the six epochs, there was a significant drift in the mean location of AKR activity over a period of 1-2 hours. The drift was predominantly in latitude at one epoch and in longitude at the other, with average drift speed V ~ 80-90 m s$^{-1}$ at the AKR emission location.

## 1. Introduction

It is now well established that auroral kilometric radiation (AKR) is generated on auroral field lines in regions of depleted electron density above the auroral zone [*Gurnett* 1974; *Kurth, et al.* 1975; *Ergun et al.* 1998]. The location of the AKR emission was first investigated by *Gurnett* [1974], who demonstrated that the time of receipt of AKR bursts observed by the Imp 6 and 8 satellites were consistent with source locations along auroral field lines near the Earth (≤ 3 Re). Gurnett also found that the probability of intense AKR bursts was closely correlated with discrete auroral arcs in the local evening region of the auroral zone. *Kurth et al*. [1975] analyzed the spin modulation of AKR bursts from the Hawkeye 1 and Imp 8 satellites to determine that the AKR bursts originated between 1 and 2 Re and were centered near 22 hours magnetic local time. *Alexander and Kaiser* [1976, 1977] used a lunar occultation technique with the lunar orbiting RAE-2 satellite to determine AKR source locations. They found that the emission extended over a wider range of Earth distances from 2 to 20 Re, including the dayside cusp region. However, these surprisingly large distances were later re-interpreted as likely due to scattering, probably by density inhomogeneities in the magnetosheath [*Alexander et al.* 1979].

[1] Department of Physics and Astronomy, University of Iowa, Iowa City, Iowa

*Gallagher and Gurnett* [1979] studied five years of electric field strength measurements of AKR bursts from the Hawkeye I and Imp 6 spacecraft and found that the AKR source location favored 65° invariant magnetic latitude and 22-24 hours magnetic local time (MLT). By measuring the phase of AKR bursts using orthogonal linear antennas and assuming the radiation is entirely R-X mode, *Huff et al*. [1988] precisely determined the direction of several strong AKR bursts at several frequencies. Assuming that the radiation arises from cyclotron resonant altitudes, they found that the inferred locations were on field lines connected to bright spots on optical auroral images taken concurrently with the DE-1 satellite. Most recently, *Schreiber et al.* [2002] used a 2-dimensional ray-tracing technique to locate two epochs of AKR burst activity observed with the Interball-2 satellite. They found that the sources were located near 22h MLT and 70° invariant latitude.

This paper describes a new technique to directly determine the location of auroral kilometric radiation bursts using very long baseline interferometry (VLBI). Although previous AKR location studies have consistently shown that AKR burst locations lie above the auroral oval and are statistically associated with auroral activity, there have been no systematic direct remote studies of the locations of AKR bursts with the ability to remotely map thousands of individual bursts with high spatial and temporal resolution. In this paper, we report the locations of over 1700 AKR bursts with uncertainties as small as 500 km in a plane perpendicular to the line of sight between the spacecraft and the source. These results are the first successful VLBI observations using an entirely space-based telescope array.

## 2. Instrumentation and Source Location Algorithm

The Cluster Wideband (WBD) receiver [*Gurnett et al.* 1997] was designed to detect low frequency radio emission with very high time (37 μs) and frequency (10 Hz) resolution. The WBD receivers are instrumented identically on all four Cluster spacecraft. Each Cluster spacecraft is equipped with two orthogonal 88 m antennas oriented in the spin plane of the spacecraft. The antennas are equipped with high-impedance preamplifiers that provide signals to the WBD receiver. Only one antenna is connected to the WBD receiver at a time, so that we could not determine polarization characteristics of the received radiation. Each WDB receiver can be tuned from 0 to 600 KHz using a variety of bandwidth and sampling options. We utilized 8-bit sampling and a 9.5 KHz instantaneous bandwidth starting at lower band edge frequencies of 125, 250, and 500 KHz. The data from each spacecraft are transmitted in real-time to NASA's Deep Space Network (DSN) downlink antennas. Each DSN station periodically time-stamps the signal using hydrogen maser time standards. The absolute accuracy of the time-stamps is nominally ±0.02 ms. This is much less than the measured differential delay uncertainty and is not included in the position uncertainty calculation.

In order to determine the AKR source position, the differential delay is measured between the waveform arrival time for each baseline between pairs of spacecraft. For baseline connecting spacecraft *i* and *j*, the differential delay is given by

$$\tau_{ij} = \frac{1}{c}\left[\left(\vec{r}-\vec{s}_j\right)-\left(\vec{r}-\vec{s}_i\right)\right], \qquad (1)$$

where $\vec{r}$ is the vector to the AKR source and $\vec{s}_{i,j}$ are vectors to spacecraft *i* and *j* respectively. This delay can be determined by cross-correlating the signals from each antenna and determining the delay offset for peak correlation.

Cross-correlation functions are calculated for each baseline by first aligning each data stream to the same UT start time. The data streams are then corrected for propagation delays to the downlink stations (the $\vec{s}_{i,j}$ vectors in equation 1). Spacecraft positions are determined from Cluster orbit files provided by the European Space Agency's Operation Center (ESOC) and are thought to be accurate to ±1 km or less. This corresponds to a timing uncertainty ±0.03 ms which is much smaller that the differential delay uncertainty, and hence is ignored.

Since individual AKR bursts are typically narrow-band with short time duration, each waveform is divided into time and frequency 'windows' before cross-correlation. Each data window consists of a 300 ms time segment and a 1 KHz frequency interval which has been filtered with a Kaiser FIR filter [*Kaiser* 1974]. For each data window, all six baseline pairs are cross-correlated using a lag window of 70 ms (±35 ms). The resulting cross-correlation function (CCF) amplitudes are searched for significant peaks using a heuristic set of threshold criteria based on tests using synthetic waveform with known noise characteristics. The resulting CCF peaks are then fitted with a Gaussian envelope function to determine the delay. Using synthetic waveform data with frequency characteristics similar to observed AKR bursts and known delays, we have determined that the Gaussian fitting procedure allows determination of the delay peak with an accuracy $\delta\tau = \pm 0.3$ ms for more than 99% of sample waveforms. We use this uncertainty in the analysis of all measured CCF peaks.

The resulting differential delays are further vetted by examining the signed sum of delays on each of the four independent baseline triangles. For an unresolved source, the signed sum of all delays on baseline triangle (*i, j, k*) must be zero within the delay uncertainty

$$\tau_{ij} + \tau_{ik} - \tau_{jk} = 0. \qquad (2)$$

This sum rule applies to all baseline triplets in an interferometric array. The locus of points corresponding to a measured differential delay on a given baseline corresponds to a hyperbolic surface whose symmetry axis is parallel to the baseline vector whose thickness is given by the delay uncertainty. If we consider three spacecraft, there are three independent delays and hence three intersecting hyperbolic surfaces whose mutual intersection is a small volume given by the simultaneous solution of equation (1). For *n* spacecraft, there are *n(n-1)/2* independent baselines. Hence, measurement of the differential delay on each baseline results in a number of intersecting delay loci whose intersection volume is the locus of all allowed source locations. For observations using all four Cluster spacecraft there are six baselines, and hence six observable delays. Therefore, the solution for the vector position (three unknowns) is over determined. This results in a robust solution algorithm, since an erroneous delay caused, for example, by a timing error at one spacecraft or downlink antenna is easily identified and rejected: For a given source location, baseline triangles containing the erroneous delay and those without a delay error cannot simultaneously satisfy the signed-summed delay equation (2).

We determine the locus of all allowed source locations using the following algorithm. First, we construct a 3-dimensional lattice of points centered on the Earth and spaced by 0.1 Re (638 km). The lattice has a dimension of 8 Re on each side, resulting in 512,000 points. Figure 1 illustrates this lattice and the paths from the event location to each Cluster spacecraft. For each set of observed delays, we calculate the propagation time from each cell in the lattice to each spacecraft. The resulting calculated differential delays are then compared with the measured values. All points which agree within the estimated uncertainty (±0.3 ms) of the observed delays for all six baselines are added to the locus of allowed solutions.

We can further constrain the locus of allowed locations by restricting the solutions to those which are located at radial distances consistent with emission generated near the electron gyro-frequency. This is appropriate for an electron-cyclotron maser emission mechanism to explain AKR bursts, as originally proposed by *Wu and Lee* [1979] and subsequently verified and modified by in situ measurements (e.g. *Louarn and Le Queau* 1996; *Ergun et al.* 1998]. We compute the surface associated with a given observed frequency equal to the electron gyro-frequency using a dipolar magnetic field approximation. The surface is defined using the following parameter mapping

$$r(f,\phi) = 9.4\,\mathrm{R}_e \cdot \left(\frac{3\cdot\sin^2\phi + 1}{f_{KHz}^2}\right)^{1/6}, \qquad (3)$$

where $f$ is the observing frequency (KHz) and $\phi$ is the geomagnetic latitude. Source locations which are within one cell of this surface are defined as the constrained solution locus.

### 2.1 Position Uncertainty Calculation

For a baseline with projected length *B* we can compute the position uncertainty $\delta x$ corresponding to a given delay uncertainty. Consider a source at distance *z* which is equidistant from both spacecraft so that the differential delay $\tau_{12} = 0$. If we displace the source in the plane perpendicular to the line of sight by distance $\delta x_\perp$ along the direction of B, the corresponding differential delay will be

$$\delta\tau = \frac{1}{c} \cdot \left\{\left[z^2 + \left(\frac{B}{2} + \delta x_\perp\right)^2\right]^{\frac{1}{2}} - \left[z^2 + \left(\frac{B}{2} - \delta x_\perp\right)^2\right]^{\frac{1}{2}}\right\} \qquad (4)$$

Assuming $\delta x << B << z$, we can easily solve for $\delta x_\perp$

$$\delta x_\perp \approx \left(\frac{z}{B}\right) \cdot c\delta\tau \qquad (5)$$

For delay uncertainty $\delta\tau = 0.3$ ms, typical maximum projected baselines $B_{max} \sim 10,000$ km – 12,000 km and source-Cluster distances $z \sim 64,000$ km – 77,000 km (Table 1), the expected position uncertainty range along the maximum projected baseline direction is $\delta x_\perp \sim 480$ km – 690 km. The differential delay is much less sensitive to a position shift along the line of sight. Assuming $\delta x_\parallel \ll B \ll z$, a calculation similar to the above, but for a source offset by B/2 in the perpendicular plane results in a position uncertainty

$$\delta x_\parallel \approx 2\left(\frac{z}{B}\right)^2 \cdot c\delta\tau \qquad (6)$$

Assuming the same parameters as in the previous uncertainty calculation, the corresponding position uncertainty range along the line of sight is $\delta x_\parallel \sim 5,120$ km – 10,700 km. Hence the derived source location locus volume will always be elongated long the line of sight with an axial ratio given by

$$\frac{\delta x_\parallel}{\delta x_\perp} \approx 2\frac{z}{B} \qquad (7)$$

Since the Cluster VLBI array consists of four spacecraft, and hence six baselines with different lengths and orientations, the cumulative locus volume will be the intersection of individual location loci. Therefore, the aggregate position uncertainty volume will be somewhat smaller than the above estimate, the exact reduction depending on the detailed array configuration.

### 2.2 Ray Bending and Propagation Speeds in the Plasmasphere

The source location algorithm used in this paper assumes a rectilinear propagation path from the AKR source location to each cluster spacecraft. There are two situations for which this assumption is likely to be incorrect. First, since AKR emission is generated in a stratified region of depleted electron density surrounded by much denser plasma, the waves cannot propagate freely. Instead, they are confined by the dense plasma cavity walls and refract upward until the density is sufficiently low to allow free propagation, with likely amplification along the path [e.g., *Gaelzer et al.* 1992; 1994]. The total distance from the source region to the region of free propagation is not well-constrained by models but is perhaps 500 km, or about of order one cell spacing. We have not attempted to correct for this effect. Hence, the observed source locations are the regions in which the radiation begins to propagate in free space rather than the cavities in which the AKR is generated.

The Earth's plasmasphere has electron densities which often exceed 1,000 particles $cm^{-3}$ near the magnetic equator [e.g., *Gallagher et al.* 2000]. The corresponding plasma frequency is comparable to or even exceeds the observing frequencies reported in this paper. Hence, propagation between the source and the spacecraft may suffer significant refractive ray bending, leading to erroneous position determinations. We have used the global core plasma of *Gallagher et al.* [2000] and a dipole magnetosphere model to calculate the expected group velocities for X-mode electromagnetic radiation and the ray bending to be expected for the present observations. For lines of sight at low magnetic latitudes ($\lambda_m < 30°$), the differential delay at the lowest observing frequency (125 KHz) can exceed the delay uncertainty of our delay measurements (±0.3 ms). However, for all six epochs reported here, the calculated differential excess delays on refracted ray paths compared with rectilinear paths is less 0.1 ms, which corresponds to one third of a resolution cell (638km). We conclude that propagation corrections are unimportant for these observations.

## 3. Observations

Since the VLBI mode requires real-time data downlinks for all four Cluster spacecraft, observations can only be scheduled when four DSN antennas are simultaneously available. In practice this limits the time available for VLBI mode to several short (1-3 hour) observing sessions per month. In this paper, we report on six observing epochs between 10 July 2002 and 22 January 2003. For the first three epochs the Cluster constellation observed AKR emission above the southern hemisphere while the latter three epochs were observed above the northern hemisphere . Each observing epoch had a total duration between one and three hours. The WBD receivers were sequenced in the 125, 250, and 500 KHz bands using a 52 – 104 – 52 second cycling time. The observing parameters are summarized in Table 1. Columns (1) and (2) list the observing epoch and UT time range of the observation. Column (3) lists the average geomagnetic activity index Kp for observation's time range. Column (4) lists the mean distance between the Earth's center and the Cluster spacecraft centroid position in Earth radii. Column (5) lists the mean magnetic latitude of the Cluster centroid. Column (6) lists the range of baseline lengths projected onto a plane perpendicular to the line of sight to the Earth. Column (7) lists the location uncertainty at the source in a plane perpendicular to the source-Cluster line of sight. Column (8) lists the corresponding uncertainty at an observing frequency of 250 KHz after the mapping the AKR source location to the auroral zone by following the magnetic field line passing through the source to 100 km altitude using the IGRF-2000 magnetic field model (Barton 1997; http://www.ngdc.noaa.gov/IAGA/wg8/igrf.html).  The footprint uncertainties for observations at frequencies 125 KHz and 500 KHz are about 25% smaller and larger respectively than those listed at 250 KHz. Figure 2 illustrates the uncertainty area for sample AKR burst at 250 KHz during epoch December 29, 2000, along with the projected uncertainty onto the auroral zone (100km altitude) using the IGRF-2000 magnetic field model.  We will refer to these projected AKR burst locations in the auroral zone as the 'magnetic footprints' of the AKR burst locations in the following discussion.

**Table 1.** Summary of Observational Parameters

| | | | VLBI Array Parameters | | | | |
|---|---|---|---|---|---|---|---|
| Epoch (1) | UT Time (hr) (2) | Kp (3) | R (4) | $\lambda_m$ (5) | Projected Baseline (km) (6) | Location Uncertainty (km) (7) | 250 KHz Footprint Uncertainty (km) (8) |
| July 10, 2002 | 08:05 – 10:05 | 1.7 | 10.6 | -62° | 2892 – 12306 | 495 – 2104 | 170 - 584 |
| July 17, 2002 | 09:05 – 11:55 | 2.3 | 12.0 | -53° | 2474 – 11923 | 578 – 2784 | 200 – 680 |
| August 19, 2002 | 17:10 – 19:25 | 3.7 | 11.9 | -72° | 2614 – 11998 | 536 – 2460 | 206 – 704 |
| November 9, 2002 | 06:25 – 08:24 | 0.8 | 10.2 | +43° | 3884 – 5402 | 1082 – 1506 | 347 – 494 |
| December 29, 2002 | 05:01 – 06:28 | 2.3 | 9.7 | +43° | 3753 – 5255 | 1058 – 1482 | 335 – 446 |
| January 22, 2003 | 00:50 – 02:48 | 3.4 | 10.6 | +42° | 4026 – 5248 | 1158 – 1510 | 366 - 462 |

**3.2 Example observation: AKR burst on July 10, 2002**

In order to illustrate the data acquisition and analysis scheme, we have chosen a sample AKR burst observation on July 10, 2002 at 08:47:06 UT. Figure 3 shows the Cluster array baseline configuration at the time of observation. The Cluster centroid was located 10.6 Re from Earth's center. Figure 3a shows the spacecraft configuration as viewed from the source projected onto the plane of the sky, while Figure 3b shows the same configuration viewed from a direction normal to the Earth-spacecraft direction. For comparison, we also show (Figure 3c, d) the corresponding baseline configurations for observations on December 29, 2002 when the Cluster constellation was above the northern hemisphere. For the epochs reported here the southern baseline configurations were more elongated, with smaller minimum baselines and larger maximum baselines than the northern configurations (Table I).

For July 10, 2002, the maximum and minimum projected baseline lengths are 11,800 km and 4,900 km respectively. From equations 5 and 6, the expected position uncertainties in the perpendicular plane range from ±495 km to ±2104 km depending on position angle. The uncertainty along the line of sight is much larger.

Figure 4a shows a 25 second dynamic spectrum (1024-point Fourier transform of the digitized waveform) in the 500 KHz - 510 KHz band. The AKR burst chosen for analysis is the boxed data window with black border, also shown in an expanded dynamic spectrum in Figure 4b. The data window was 300 ms long and was bandpass filtered from 506 KHz - 507 KHz. The boxes with white borders are additional data windows for which burst locations were also determined. The filtered waveforms from all four spacecraft are shown in Figure 5.

The cross-correlation functions of the waveforms on all six baselines are shown is Figure 6. The delay peaks were fitted with Gaussian functions using a least-squares method to obtain the corresponding delays. The locus of valid position solutions is shown in Figure 7. Figure 7a shows the unconstrained locus of locations from the perspective of the Cluster spacecraft centroid, while Figure 7b shows the same solutions viewed obliquely. The position locus volume is consistent with the estimate given above, viz. approximately 3 x 5 cells in the perpendicular plane, but several Earth radii in the line of sight direction. It is clear that the delay mapping technique is much more sensitive to position in the plane orthogonal to the line of sight between the source and the spacecraft than along it. Figure 7c shows the constrained locations using the further constraint that the points lie in the surface in which the observing frequency is equal to the electron gyrofrequency. We have superimposed a statistical auroral oval derived from the model of *Feldstein and Starkov* [1967] for a moderate geomagnetic activity index. We have also added magnetic field lines passing through the calculated locus of locations to the Earth's surface to the corresponding magnetic footprint location using the IGRF-2000 geomagnetic field model as described above.

**4. Locations of Auroral Kilometric Radiation Bursts: Statistical Description**

In order to determine the statistical properties of AKR bursts locations, we performed a systematic search using all eleven hours of 4-spacecraft waveforms over the six epochs listed in Table 1. For each epoch, every 300 ms x 1 KHz data window was cross-correlated on all six baselines after shifting the data streams by the appropriate downlink propagation delay. The resulting cross-correlation functions (CCF) were searched for significant delays peaks. The delays were then filtered using the signed-sum criterion (equation 2) on all baseline triplets. All delay sums which did sum to zero within the delay uncertainty on all baseline triplets were rejected.

We determined locations for over 1700 individual AKR bursts. In Table 2 we list the statistical properties of the locations for each of the six epochs. Column (2) lists the observed hemisphere (northern or southern). Column (3) lists the number of individual AKR burst locations determined for that epoch. Column (4) lists the median local time for the locations, and column (5) lists the median invariant magnetic latitude. Column (6) lists the approximate overall size of the burst locations for each epoch mapped to an altitude of 100 km (magnetic footprint). Figure 8 shows the distribution of AKR burst locations at each epoch in corrected geomagnetic coordinates. Each plotted point results from mapping the mean location of an individual AKR burst to an altitude of 100 km along a magnetic filed line passing through the burst location using the IGRF-2000 magnetic field model (Barton 1997). The colors encode the observed frequency band as described in the figure caption.

Several trends are evident in these maps: (1) The mean magnetic local time (MLT) of bursts in the southern hemisphere favors the morning quadrant, whereas the northern hemisphere locations favor the evening sector. The mean MLT for the three southern hemisphere epochs is $0.5^h \pm 1.4^h$ while for the northern hemisphere it is $19.1^h \pm 1.4^h$ , (2) The southern hemisphere mean invariant latitude is located somewhat poleward of the center of the statistical auroral oval (mean value -73.9° ± 1.6°) whereas in the northern hemisphere, the mean invariant latitude is much closer to the center of the auroral oval (mean value 70.0° ± 1.2°); (3) Over a 1-3 hour timescale, individual AKR burst locations are confined to a relatively small fraction of the auroral oval, with a typical active region 'footprint' having dimensions between 1000 and 4000 km when mapped into the auroral zone. We have not yet investigated whether there is a statistically significant difference in the source location with frequency. This question will be addressed in a future paper.

**Table 2.** AKR Burst Locations

AKR Burst Location Statistics

| Epoch (1) | N/S (2) | Nr. (3) | <MLT (hr)> (4) | <$\lambda_m$> (5) | CGM Footprint Size (km x km) (6) |
|---|---|---|---|---|---|
| 10 Jul 2002 | S | 319 | $0.8^h$ | -75.1° | 2100 x 1400 |
| 17 Jul 2002 | S | 81 | $2.1^h$ | -71.7° | 2340 x 940 |
| 19 Aug 2002 | S | 171 | $22.7^h$ | -75.0° | 1870 x 1640 |
| 9 Nov 2002 | N | 568 | $20.7^h$ | +68.4° | 2340 x 1400 |
| 29 Dec 2002 | N | 221 | $19.3^h$ | +70.2° | 3280 x 1170 |
| 22 Jan 2003 | N | 372 | $18.2^h$ | +71.0° | 2050 x 1640 |

## 5. Discussion

### 5.1 Comparison with Previous AKR Burst Location Studies

Previous studies of AKR source locations showed a strong tendency for AKR locations to favor the evening sector and invariant magnetic latitudes within the auroral zone. Early source location papers [e.g., *Kurth et al.* 1975, *Gallagher and Gurnett* 1979; *Kaiser and Alexander* 1977] all found a preference for the pre-midnight hours between 21-24 LT. However, many of these early results were obtained with satellites favoring northern hemisphere observations. For example, the Hawkeye I satellite used by *Kurth et al*. [1975] was in a highly elliptical orbit with apogee above the northern hemisphere, while *Kaiser and Alexander* [1977] utilized data from the Imp-6 satellite, which was in an inclined orbit with apogee always above +17° invariant latitude, favoring northern hemisphere detections. The present results for the three northern hemisphere epochs are in good agreement with these earlier results. We are not aware of any previous AKR burst location investigations that have been able to study the distribution of AKR sources in the southern hemisphere.

We also find that the median invariant magnetic latitude of AKR emission (+70.0° northern hemisphere , -73.9° southern hemisphere) is somewhat more poleward than previous studies. For example *Gallagher and Gurnett* [1979], who analyzed five years of AKR emission from the Hawkeye I and Imp 6 satellites, found that the emission locations had a invariant latitudes between +57° and +69°, although this result is strongly biased toward the northern hemisphere as discussed above. The present results suggest although the northern hemisphere latitudes are consistent with past observations and within the auroral zone, southern hemisphere AKR emission originates preferentially near the poleward edge of the auroral oval. Unfortunately, auroral imaging observations were not available during any of the epochs reported here, so that we cannot determine the latitudinal extent of the auroral oval during our observations. However, we can estimate the latitudinal extent of the auroral zone using the model of *Feldstein and Starkov* [1967]. For all three southern epochs geomagnetic activity was low to moderate, indicating that poleward edge of the auroral zone lies between 70° - 72° near local midnight, and hence that our reported locations are within or poleward of this range.

The systematic differences in mean MLT of AKR sources in the northern and southern hemispheres may be partly a consequence of the definition of MLT used in the IGRF-2000 coordinate system. In this system, MLT is defined as the time difference between the UT time of observation and the UT time at local midnight (cf. http://nssdc.gsfc.nasa.gov/space/cgm/cgmm_des.html). *Frank and Sigwarth* [2003] recently reported the first simultaneous images of northern and southern aurora using the VIS camera on the Polar spacecraft. Using the same definition of MLT, they found a magnetic local time difference of about 40 min for several bright features on opposite hemispheres, with the southern bright feature located eastward (toward morning) with respect to the northern feature. *Burns et al.* [1990] had previously reported a similar result, finding that comparison of three conjugate auroral events from South Pole Station and the Viking spacecraft camera in the northern hemisphere showed southern features which were located up to 1.6 hours east of the conjugate northern features.

**5.2 Multiple, Spatially Separated Sources**

If multiple sources of AKR bursts separated by more than one resolution cell (638 km) contribute to the signal in a single data window, there will be multiple delays peaks, or perhaps a continuous range of delay peaks. Since our algorithm selects a single isolated peak in each CCF, it will only succeed if there is a single intense AKR burst at a given time or a dominant burst in a spatially separated group of bursts. For multiple spatially separated sources with similar intensities, no solution will be found. We find that the fraction of all data windows which result in valid position solutions is very small at all epochs, even in spectra with substantial AKR burst activity. A typical example is the 10 July 2002 epoch, which had almost continuous AKR burst activity at all frequencies for the entire two hour observation. There were only 319 burst location solutions, while the total number of data windows was over 250,000. This means that only 2% of all data windows produced valid location solutions, whereas more than 50% had detectable AKR emission. Figure 4a illustrates this nicely: During the entire 25 sec interval, only 23 valid burst locations were determined (white boxes), while more than 50% of the data windows (~400) had sufficiently intense AKR emission for cross-correlation analysis. We believe that this may be a result of multiple, spatially separated sources dominating the total AKR emission in each data window. In this case, there will be multiple delay peaks in the cross-correlation function and the 'single peak' CCF criterion (section 2) will fail to find a solution. We are currently investigating whether it will be possible to modify the location algorithm to account for multiple spatially separated sources.

**5.3 AKR Location Variations on Short Timescales**

Figure 8 shows that within each epoch there is a large variation in AKR source locations, both in MLT and invariant latitude. Some of this variation is a result of the uncertainty of individual source location, as discussed in section 2.1. For the sample burst on 10 July 2002 at 08:47:06 UT, the constrained solution (figure 7c) has an approximate spread of 400 x 600 km at the magnetic footpoint. However, this is much smaller than the overall spread in source locations for this epoch (Table 2), indicating that the overall extent of locations is due to a range of individual burst locations which could drift systematically on a short timescale. We have analyzed the mean source locations as a function of time for the six epochs reported in Table 2. For four epochs we did not find any systematic motion, but for two epochs (August 19, 2002, and January 22, 2003) there were clear trends. Figure 9 shows the variation in CGM coordinates of AKR footpoint locations for these epochs as a function of UT. In figure 9a (August 19, 2002), there is a clear drift in invariant latitude, about $5°$ $hr^{-1}$, corresponding to a poleward speed 80 m $s^{-1}$ at the height of the AKR source. Figure 9b (January 22, 2003) shows a drift in MLT, about 2.2 hr $hr^{-1}$, or 90 m $s^{-1}$ at the height of the AKR source.

**6. Conclusions**

1. We have used differential delay triangulation from multiple Cluster spacecraft to determine the locations of more than 1700 AKR bursts above the northern and southern polar regions during six epochs between July 10, 2002 and January 22, 2003. The delays are determined by standard interferometric cross-correlation of the received waveforms. These observation constitute the first very long baseline interferometer (VLBI) observations using a spaced-based array of multiple antennas.

2. The uncertainty in source location for a typical projected baseline length of 10,000 km is ±600 km in a plane perpendicular to the line of sight between the spacecraft and the source. The position uncertainty is much larger (10,000 km or more) along the line of sight. We constrained the line of sight source position by assuming that the radiation mechanism was an electron cyclotron maser so that the source region was at a height such that the electron gyrofrequency matched the observing frequency. The position uncertainty is dominated by the width of the cross-correlation delay peak uncertainty (typically ±0.3 ms), rather than systematic uncertainties in instrumental clocks or propagation time. The delay peak width is, in turn, a result of the narrow-band nature of the AKR bursts.

3. For all three observing epochs with the Cluster constellation over the northern hemisphere, the AKR source locations favored the local evening sector, with mean magnetic local times varying from 18.2 to 20.7 hours. The corresponding invariant magnetic latitudes varied from 68.4° to 71.0°. These results indicate that the northern AKR locations are located along magnetic field lines connecting the statistical auroral oval and are in good agreement with previous AKR location studies.

4. For the three observing epochs over the southern hemisphere, the geomagnetic coordinates of AKR location footprints were significantly shifted with respect to the northern hemisphere locations. The source locations had mean magnetic local times close to midnight or in the morning sector (from 22.7 to 2.1 hours) and invariant magnetic latitudes somewhat poleward of the statistical auroral oval (mean values from -71.7° to -75.1°).

5. On timescales of 1-3 hours the locus of AKR footprint locations at the auroral zone had a characteristic spatial scale between 1000 - 4000 km, significantly larger than the positional uncertainty of an individual AKR burst location magnetic footprint, but a small fraction of the auroral oval. This indicates that on this timescale, conditions for suitable generation of AKR emission are found on only a fraction of all magnetic field lines crossing through the auroral oval.

6. For two of the six epochs, there was a significant drift in the mean location of AKR activity over a period of 1-2 hours. The drift was predominantly in latitude at one epoch and in longitude at the other, with average drift speed V ~ 80-90 m s$^{-1}$ at the AKR emission location.

7. Since the VLBI delay-mapping technique selects isolated peaks in the baseline cross-correlation functions, it rejects events for which multiple, spatially separated AKR regions are simultaneously emitting with similar intensities. Hence the statistical properties of the source locations reported in this paper refer only to bright, isolated AKR sources, rather than multiple source events, for which the location statistics may be rather different.

**Acknowledgements.** This research is supported by NASA through contract NAS5-30730 and grant NAG5-9974 from Goddard Space Flight Center. The authors are grateful to Justin Cook for assistance with the data analysis.

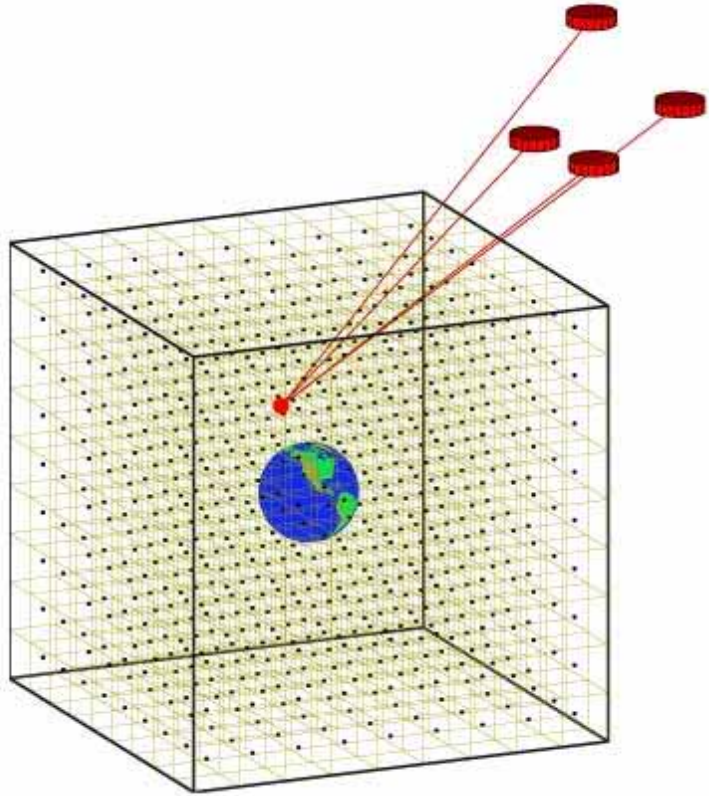

**Figure 1.** Illustration of the position search algorithm. A uniform 3-d grid of points is constructed centered on the Earth with spacing 0.1 Re and dimension 8 Re on each side (512,000 grid points). A coarser grid size of 0.5 Re is shown here for clarity. The propagation time to each satellite is computed from each grid point. Differential delays are then compared with observed delays, as measured by cross-correlating the waveforms from each pair of spacecraft.

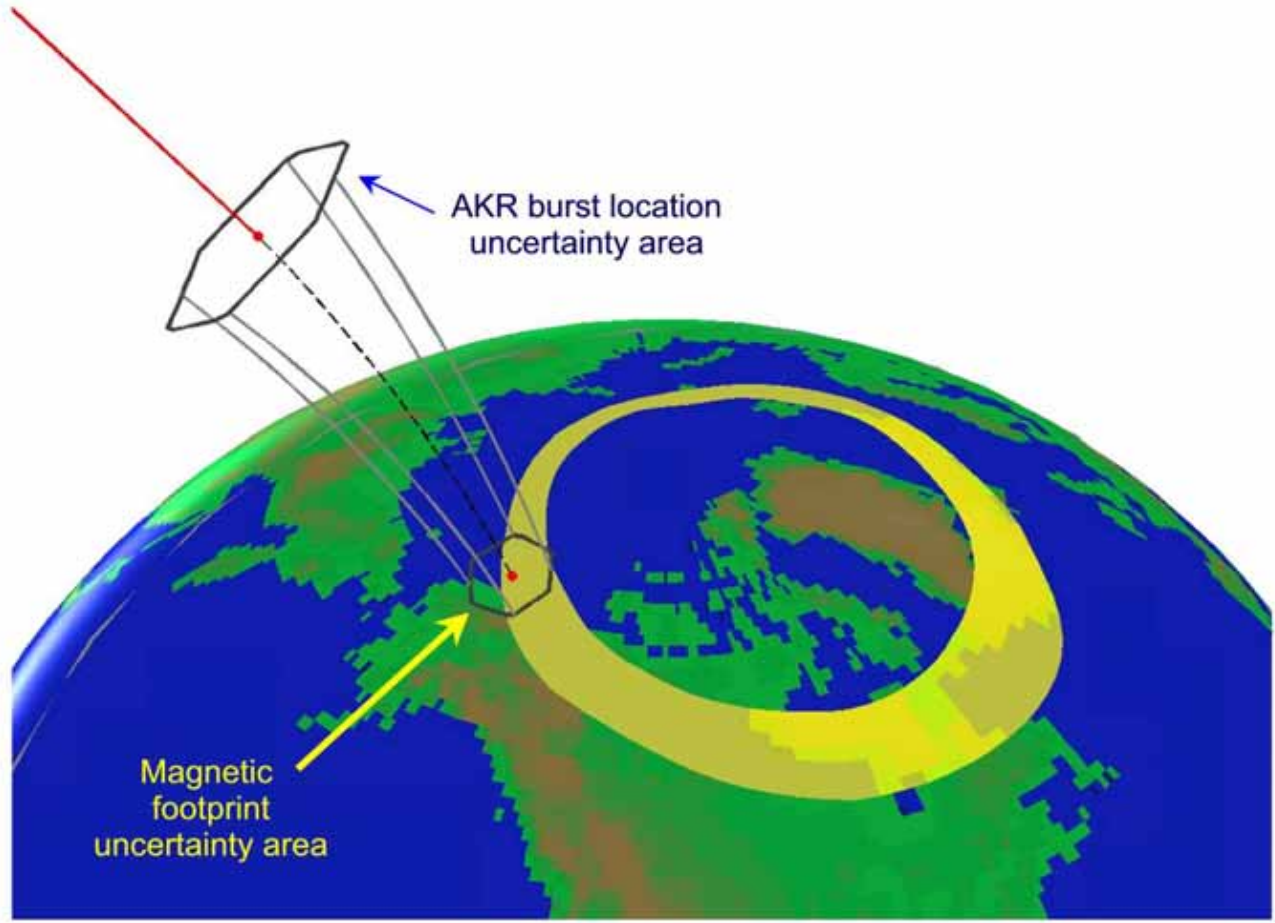

**Figure 2.** The uncertainty area for a sample 250 KHz AKR burst location at epoch December 29, 2002 calculated using equation (5) and assuming that the burst arises from a height such that the electron gyro-frequency is equal to the observing frequency. The uncertainty area is mapped onto a plane normal to the source-Cluster line of sight at the source altitude. This area is projected onto the auroral zone (100 km altitude) by following magnetic field lines (gray lines) using the IGRF-2000 model. In the text, the mapped uncertainty area in denoted the 'magnetic footprint' of that AKR burst location.

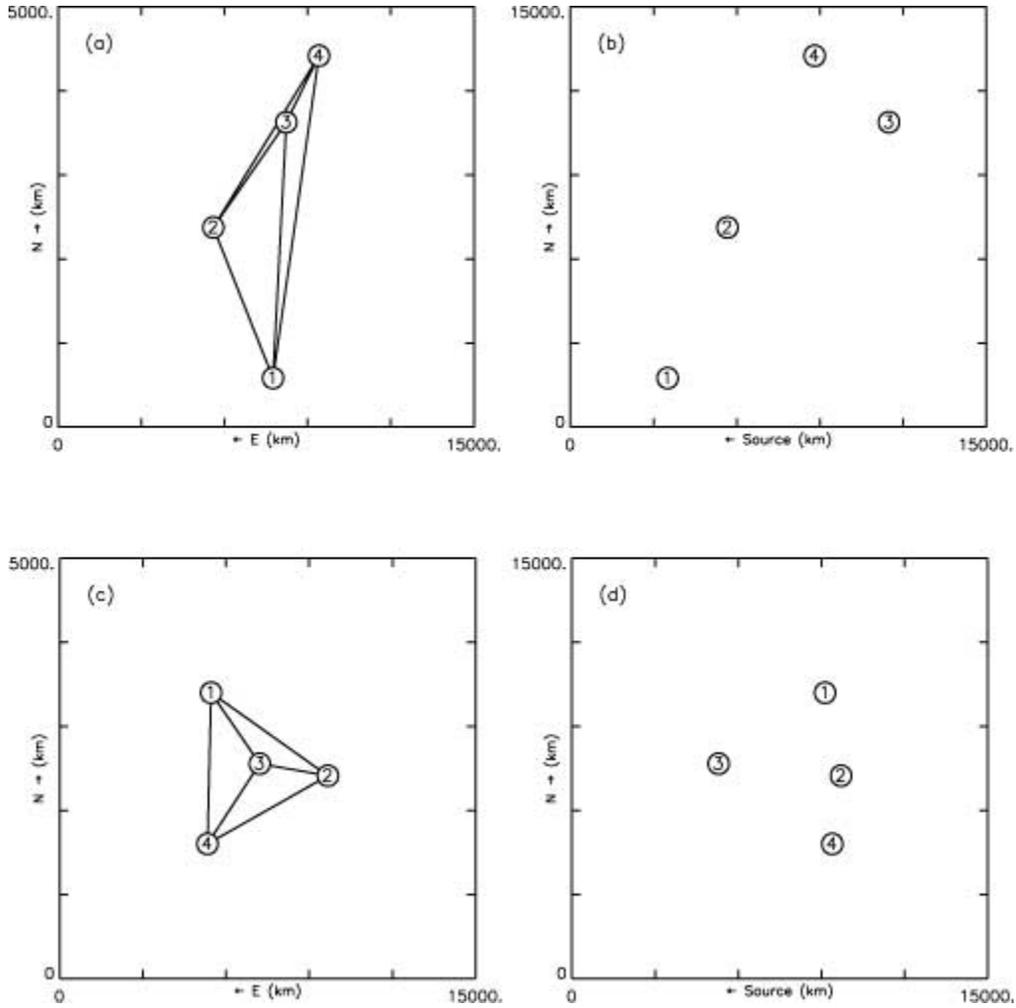

**Figure 3.** (*(a)* Projected Cluster spacecraft locations and baselines as viewed from AKR source region on July 10, 2002 at 08:46 UT. *(b)* Same as *(a),* but as viewed from a direction normal to the source-spacecraft direction; *(c)* Same as *(a),* but for December 29, 2002; *(d)* Same as *(b),* but for December 29, 2002.

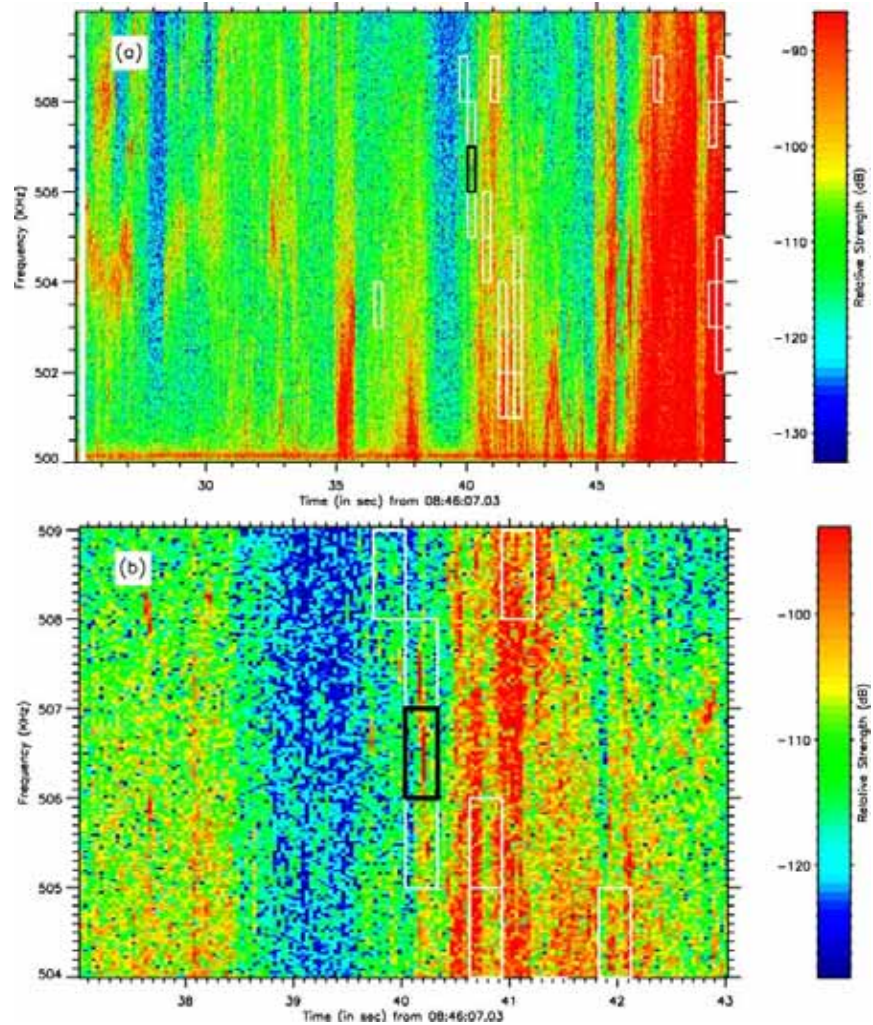

**Figure 4.** (a) Dynamic spectrum of AKR bursts on 10 July 2002 from 08:46:32 to 08:46:50 UT. The relative intensity of the bursts is indicated by color on the logarithmic scale at right. The light white boxes indicate cross-correlation 'hits' which satisfied the cross-correlation peak and delay sum constraints discussed in the text. The black-white checkered box is the filtered waveform shown in Figure 4 and whose cross-correlations are shown in Figure 5.

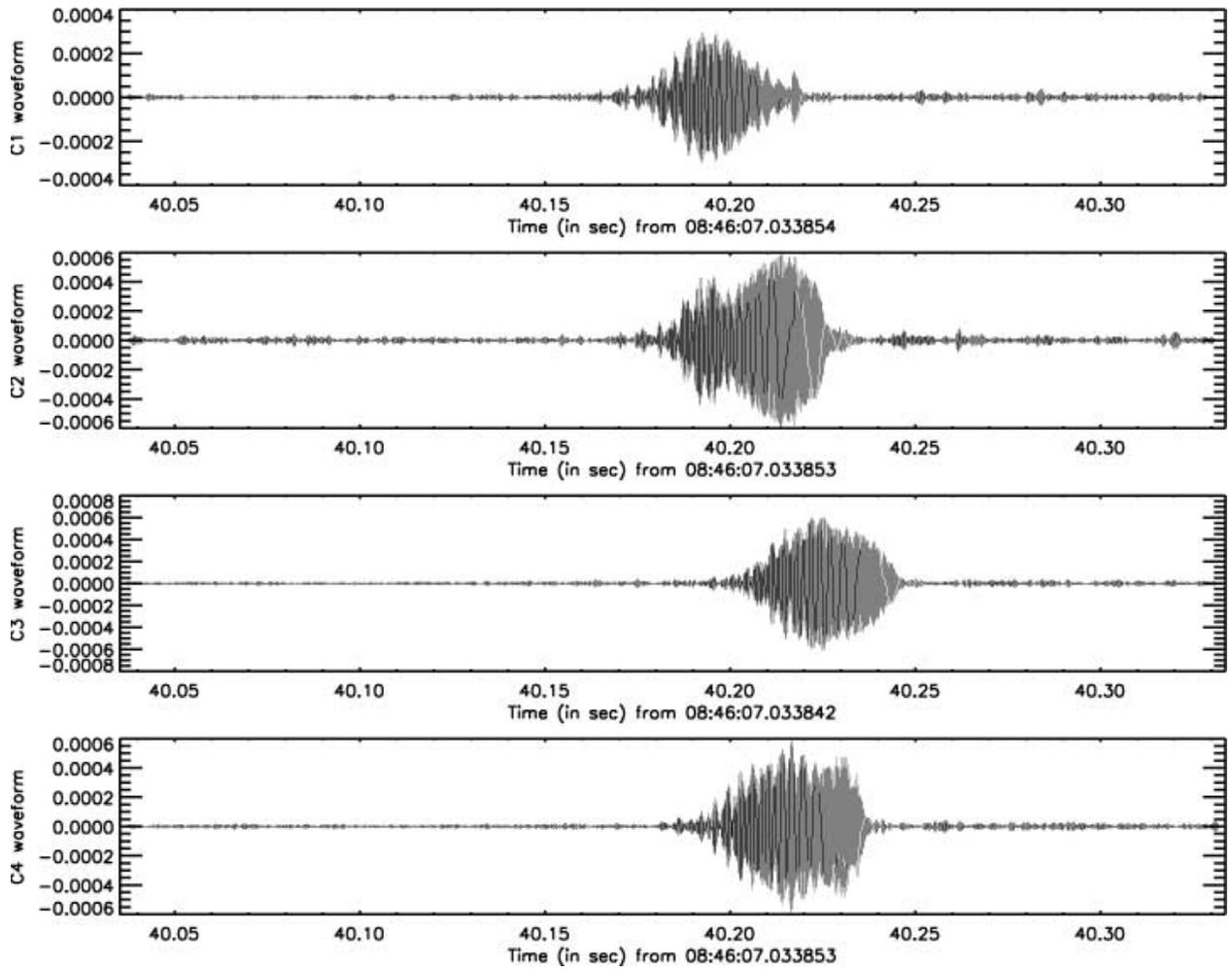

**Figure 5.** Filtered AKR burst waveforms received on cluster spacecraft C1, C2, C3, and C4. The waveforms are for the checkered box illustrated in Figure 4 (300 ms time interval, 1 KHz bandpass filtered from 506 to 507 KHz).

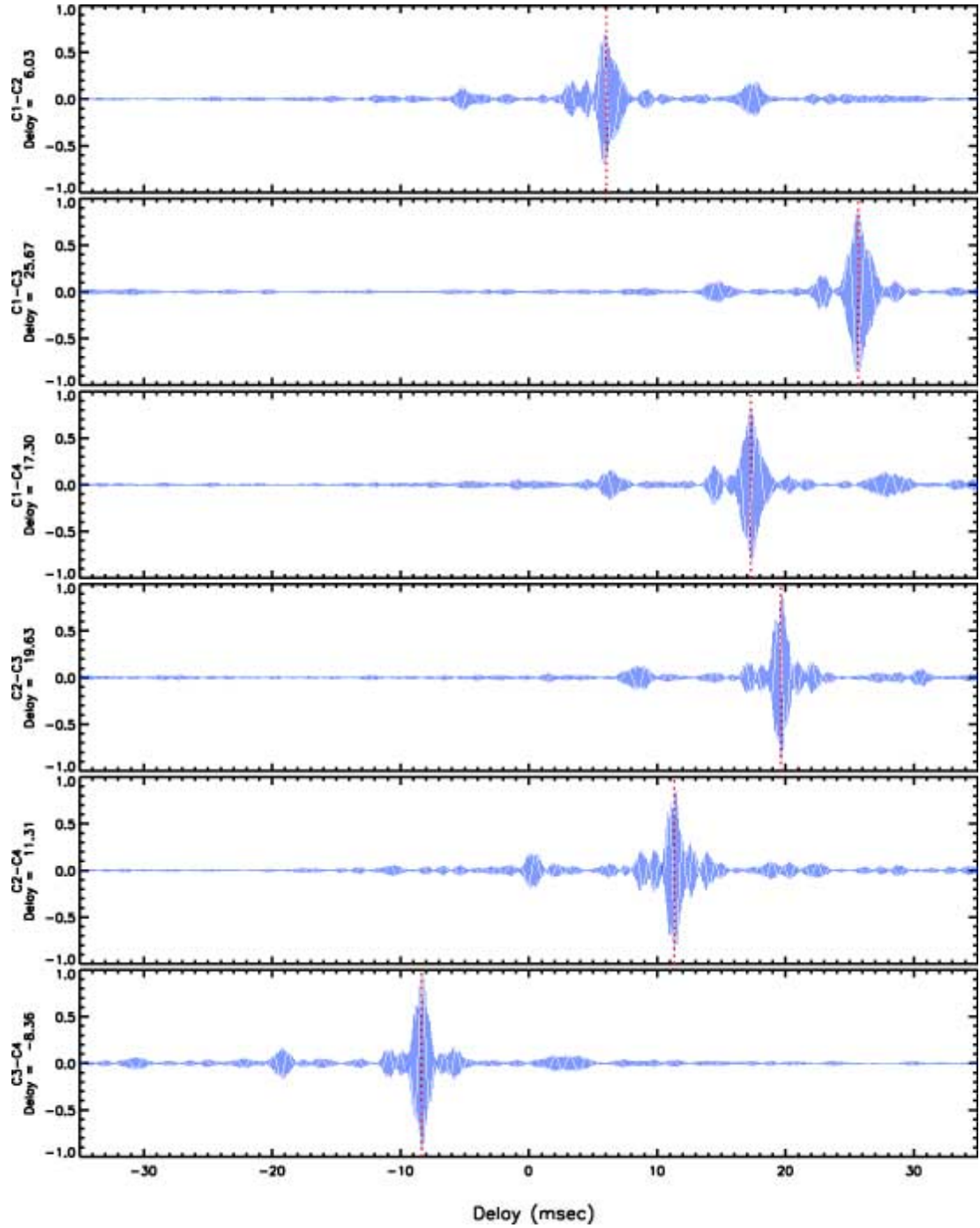

**Figure 6.** Cross-correlation functions of WBD waveforms received from all four Cluster spacecraft on 10 July 2002 starting at 08:46:53.04 UT. The waveforms were filtered in the time and frequency domain using windows of 300 ms and 1 KHz (506 KHz to 507 KHz) respectively.

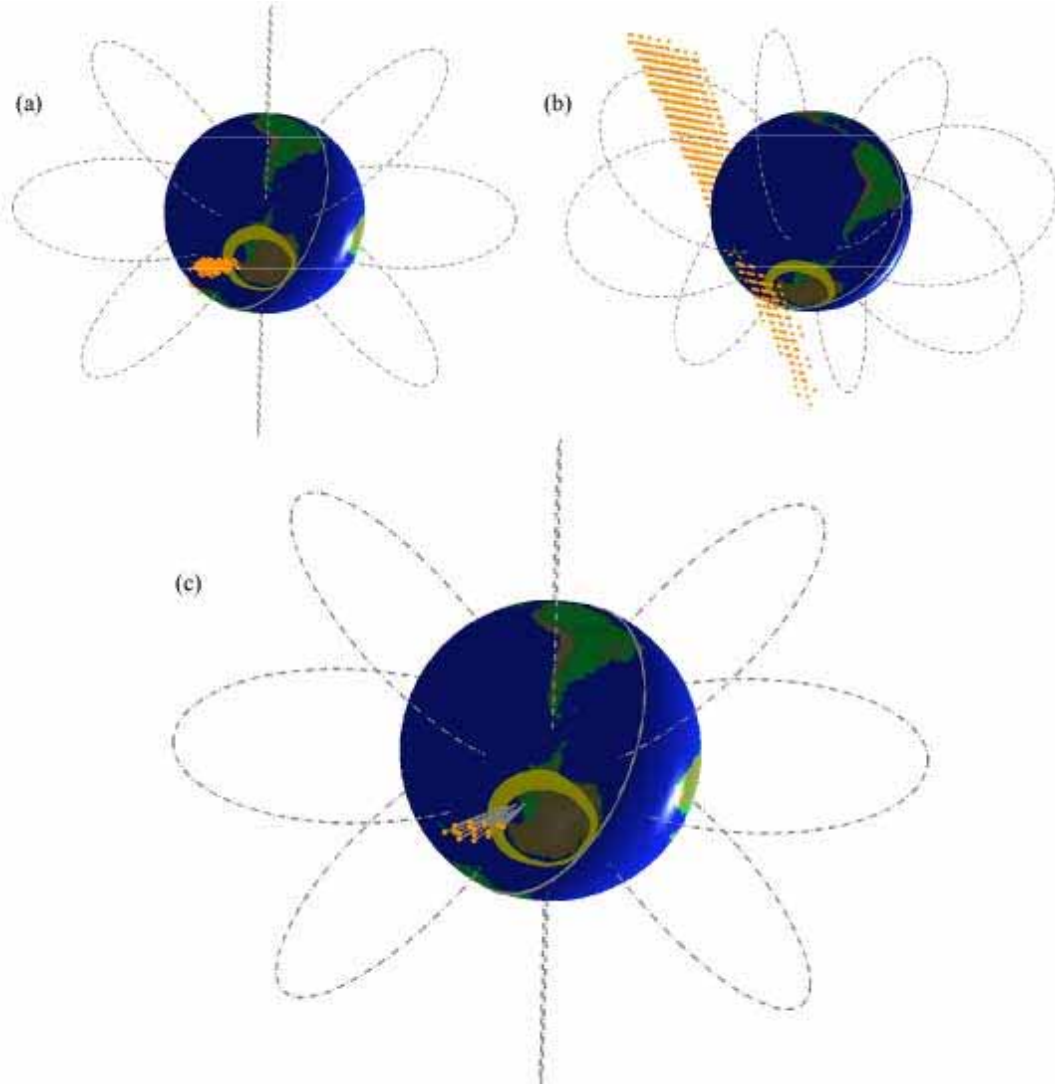

**Figure 7.** (a) Locus of all derived positions (orange dots) consistent with the measured delays for the AKR burst observed on 10 July 2002 at 08:47:06 UT. (b) Same as (a), but viewed from an oblique angle, showing the elongated structure of the locus of allowed solutions. (c), Same as (a), but constraining the allowed positions to lie within 3% of the expected radius consistent with the gyro-frequency of the observed emission, calculated using a dipolar magnetic field model. The light gray lines are the magnetic field line passing through the allowed positions to the Earth's surface. The light yellow oval is a model auroral zone oval based on a moderate activity level [*Felstein and Starkov* 1967].

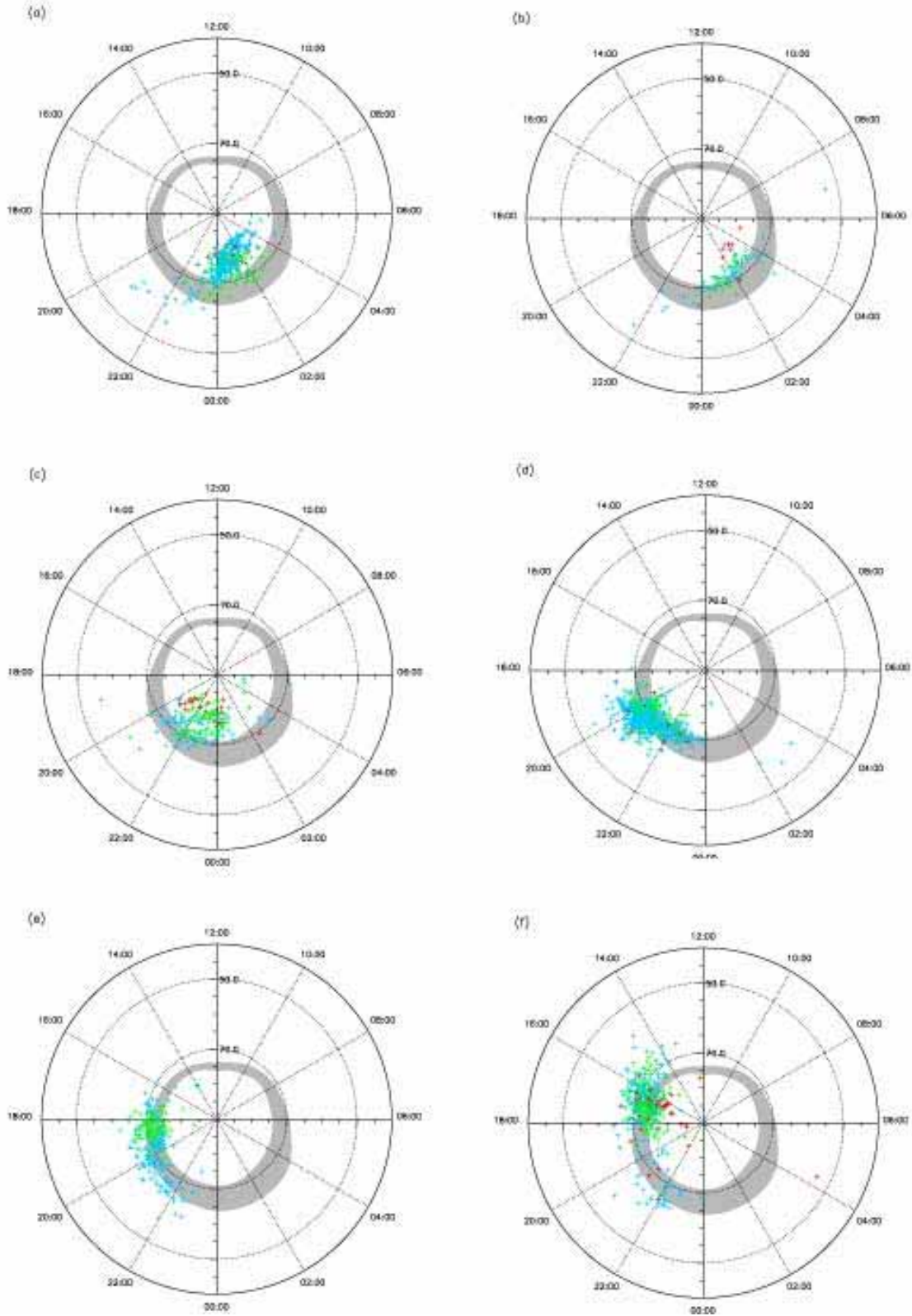

**Figure 8.** AKR burst positions observed at each epoch listed in Table II as a function of corrected geomagnetic coordinates. A statistical auroral oval (gray) is superposed for reference, based on the model of *Feldstein and Starkov* [1967] for a moderate Kp index. Each cross is the centroid of solutions for an individual AKR burst at 125 - 135 KHz band (red crosses), 250 - 260 KHz band (green crosses), and 500 - 510 KHz band (blue crosses). Epoch labels are: *(a)* 10 July 2002 (S hemisphere), *(b)* 17 July 2002 (S hemisphere), *(c)* 19 August 2002 (S hemisphere), *(d)* 9 November 2002 (N hemisphere), *(e)* 29 December 2002 (N hemisphere), and *(f)* 22 January 2003 (N hemisphere).

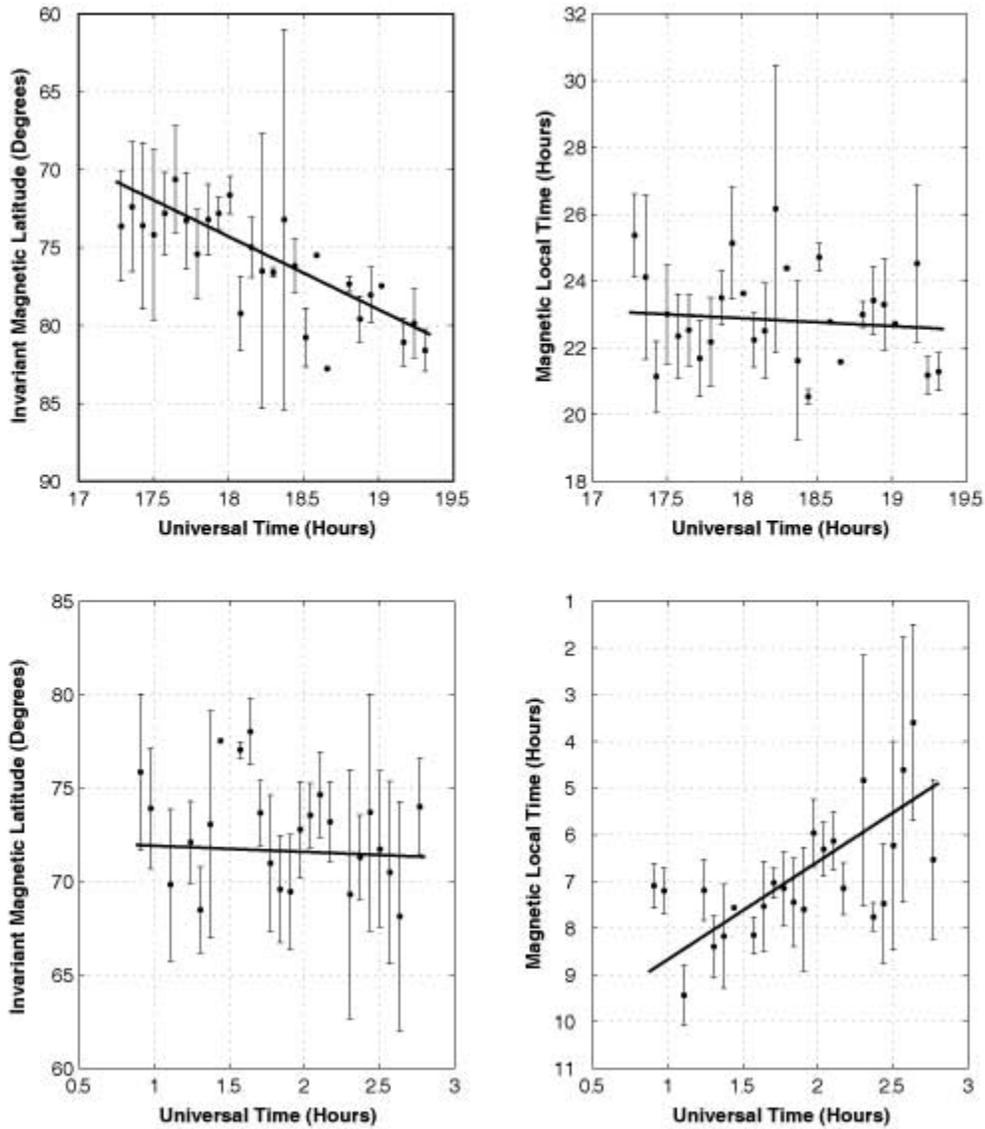

**Figure 9.** (Top panel) AKR locations versus UT time in CGM coordinates for observations at epoch August 19, 2002. The locations have been binned using a 5 minute sampling window. Note the migration in invariant magnetic latitude. (Bottom panel) Same as above, but for observing epoch January 22, 2003. For this epoch, the location migration was in magnetic local time.